# HORSESHOES IN MULTIDIMENSIONAL SCALING AND LOCAL KERNEL METHODS

By Persi Diaconis,[1] Sharad Goel[2] and Susan Holmes[3]

*Stanford University, Yahoo! Research and Stanford University*

Classical multidimensional scaling (MDS) is a method for visualizing high-dimensional point clouds by mapping to low-dimensional Euclidean space. This mapping is defined in terms of eigenfunctions of a matrix of interpoint dissimilarities. In this paper we analyze in detail multidimensional scaling applied to a specific dataset: the 2005 United States House of Representatives roll call votes. Certain MDS and kernel projections output "horseshoes" that are characteristic of dimensionality reduction techniques. We show that, in general, a latent ordering of the data gives rise to these patterns when one only has *local* information. That is, when only the interpoint distances for nearby points are known accurately. Our results provide a rigorous set of results and insight into manifold learning in the special case where the manifold is a curve.

**1. Introduction.** Classical multidimensional scaling is a widely used technique for dimensionality reduction in complex data sets, a central problem in pattern recognition and machine learning. In this paper we carefully analyze the output of MDS applied to the 2005 United States House of Representatives roll call votes [Office of the Clerk—U.S. House of Representatives (2005)]. The results we find seem stable over recent years. The resultant 3-dimensional mapping of legislators shows "horseshoes" that are characteristic of a number of dimensionality reduction techniques, including principal components analysis and correspondence analysis. These patterns are heuristically attributed to a latent ordering of the data, for example, the ranking of politicians within a left-right spectrum. Our work lends insight into this heuristic, and we present a rigorous analysis of the "horseshoe phenomenon."

Received June 2007; revised January 2008.

[1]This work was part of a project funded by the French ANR under a Chaire d'Excellence at the University of Nice Sophia-Antipolis.

[2]Supported in part by DARPA Grant HR 0011-04-1-0025.

[3]Supported in part by NSF Grant DMS-02-41246.

*Key words and phrases.* Horseshoes, multidimensional scaling, dimensionality reduction, principal components analysis, kernel methods.

Seriation in archaeology was the main motivation behind D. Kendall's discovery of this phenomenon [Kendall (1970)]. Ordination techniques are part of the ecologists' standard toolbox [ter Braak (1985, 1987), Wartenberg, Ferson and Rohlf (1987)]. There are hundreds of examples of horseshoes occurring in real statistical applications. For instance, Dufrene and Legendre (1991) found that when they analyzed the available potential ecological factors scored in the form of presence/absence in 10 km side squares in Belgium there was a strong underlying gradient in the data set which induced "an extraordinary horseshoe effect." This gradient followed closely the altitude component. Mike Palmer has a wonderful "ordination website" where he shows an example of a contingency table crossing species counts in different locations around Boomer Lake [Palmer (2008)]. He shows a horseshoe effect where the gradient is the distance to the water (Palmer). Psychologists encountered the same phenomenon and call it the Guttman effect after Guttman (1968). Standard texts such as Mardia, Kent and Bibby (1979), page 412, claim horseshoes result from ordered data in which only local interpoint distances can be estimated accurately. The mathematical analysis we provide shows that by using the exponential kernel, any distance can be downweighted for points that are far apart and also provide such horseshoes.

Methods for accounting for [ter Braak and Prentice (1988)], or removing gradients [Hill and Gauch (1980)], that is, detrending the axes, are standard in the analysis of MDS with chisquare distances, known as correspondence analysis.

Some mathematical insights into the horseshoe phenomenon have been proposed [Podani and Miklos (2002), Iwatsubo (1984)].

The paper is structured as follows: In Section 1.1 we describe our data set and briefly discuss the output of MDS applied to these data. Section 1.2 describes the MDS method in detail. Section 2 states our main assumption—that legislators can be isometrically mapped into an interval—and presents a simple model for voting that is consistent with this metric requirement. In Section 3 we analyze the model and present the main results of the paper. Section 4 connects the model back to the data. The proofs of the theoretical results from Section 3 are presented in the Appendix.

1.1. *The voting data.* We apply multidimensional scaling to data generated by members of the 2005 United States House of Representatives, with similarity between legislators defined via roll call votes (Office of the Clerk—U.S. House of Representatives). A full House consists of 435 members, and in 2005 there were 671 roll calls. The first two roll calls were a call of the House by States and the election of the Speaker, and so were excluded from our analysis. Hence, the data can be ordered into a $435 \times 669$ matrix $D = (d_{ij})$ with $d_{ij} \in \{1/2, -1/2, 0\}$ indicating, respectively, a vote of "yea," "nay," or "not voting" by Representative $i$ on roll call $j$. (Technically, a representative can vote "present," but for purposes of our analysis this was treated as equivalent to "not voting.") We further restricted our analysis to the 401 Representatives that voted on at least 90% of the roll calls (220 Republicans, 180 Democrats and 1 Independent), leading to a $401 \times 669$ matrix $V$ of voting data. This step removed, for example, the Speaker of House Dennis Hastert (R-IL) who by custom votes only when his vote would be decisive, and Robert T. Matsui (D-CA) who passed away at the start the term.

As a first step, we define an empirical distance between legislators as

$$(1.1) \qquad \hat{d}(l_i, l_j) = \frac{1}{669} \sum_{k=1}^{669} |v_{ik} - v_{jk}|.$$

Roughly, $\hat{d}(l_i, l_j)$ is the percentage of roll calls on which legislators $l_i$ and $l_j$ disagreed. This interpretation would be exact if not for the possibility of "not voting." In Section 2 we give some theoretical justification for this choice of distance, but it is nonetheless a natural metric on these data.

Now, it is reasonable that the empirical distance above captures the similarity of nearby legislators. To reflect the fact that $\hat{d}$ is most meaningful at small scales, we define the proximity

$$P(i, j) = 1 - \exp(-\hat{d}(l_i, l_j)).$$

Then $P(i, j) \approx \hat{d}(l_i, l_j)$ for $d(l_i, l_j) \ll 1$ and $P(i, j)$ is not as sensitive to noise around relatively large values of $\hat{d}(l_i, l_j)$. This localization is a common feature of dimensionality reduction algorithms, for example, eigenmap [Niyogi (2003)], isomap [Tenenbaum, de Silva and Langford (2000)], local linear embedding [Roweis and Saul (2000)] and kernel PCA [Schölkopf, Smola and Muller (1998)].

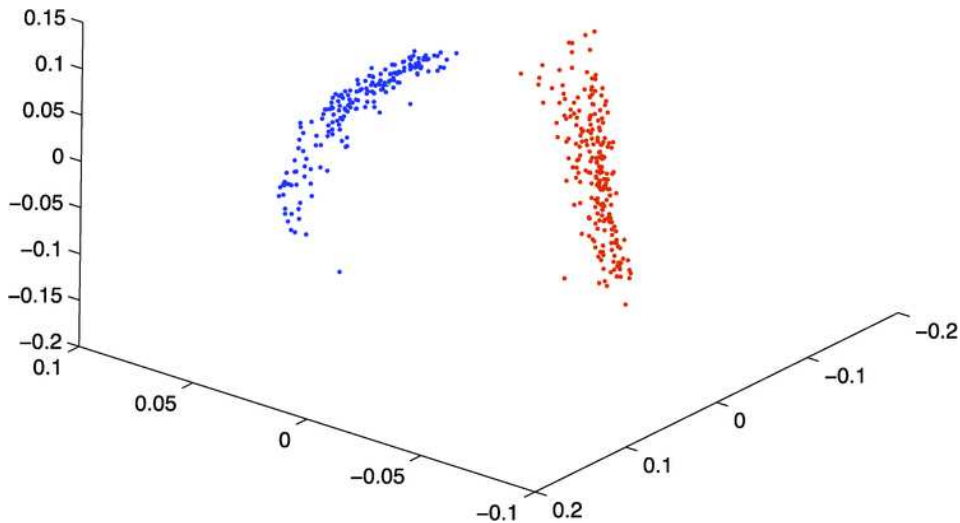

FIG. 1.  *3-Dimensional MDS output of legislators based on the 2005 U.S. House roll call votes. Color has been added to indicate the party affiliation of each Representative.*

We apply MDS by double centering the squared distances built from the dissimilarity matrix $P$ and plotting the first three eigenfunctions weighted by their eigenvalues (see Section 1.2 for details). Figure 1 shows the results of the 3-dimensional MDS mapping. The most striking feature of the mapping is that the data separate into "twin horseshoes." We have added color to indicate the political party affiliation of each Representative (blue for Democrat, red for Republican and green for the lone independent—Rep. Bernie Sanders of Vermont). The output from MDS is qualitatively similar to that obtained from other dimensionality reduction techniques, such as principal components analysis applied directly to the voting matrix $V$.

In Sections 2 and 3 we build and analyze a model for the data in an effort to understand and interpret these pictures. Roughly, our theory predicts that the Democrats, for example, are ordered along the blue curve in correspondence to their political ideology, that is, how far they lean to the left. In Section 4 we discuss connections between the theory and the data. In particular, we explain why in the data legislators at the political extremes are not quite at the tips of the projected curves, but rather are positioned slightly toward the center.

1.2. *Multidimensional scaling.* Multidimensional Scaling (MDS) is a widely used technique for approximating the interpoint distances, or dissimilarities, of points in a high-dimensional space by actual distances between points in a low-dimensional Euclidean space. See Young and Householder (1938) and Torgerson (1952) for early, clear references, Shepard (1962) for extensions from distances to ranked similarities, and Mardia, Kent and Bibby (1979), Cox and Cox (2000) and Borg and Groenen (1997) for useful textbook accounts. In our setting, applying the usual centering operations of MDS to the proximities we use as data lead to surprising numerical coincidences: the eigenfunctions of the centered matrices are remarkably close to the eigenfunctions of the original proximity matrix. The development below unravels this finding, and describes the multidimensional scaling procedure in detail.

*Euclidean points*: If $x_1, x_2, \ldots, x_n \in \mathbb{R}^p$, let

$$d_{i,j} = \sqrt{(x_i^1 - x_j^1)^2 + \cdots + (x_i^p - x_j^p)^2}$$

be the interpoint distance matrix. Schoenberg [Schoenberg (1935)] characterized distance matrices and gave an algorithmic solution for finding the points given the distances (see below). Albouy (2004) discusses the history of this problem, tracing it back to Borchardt (1866). Of course, the points can only be reconstructed up to translation and rotation, thus, we assume $\sum_{i=1}^{n} x_i^k = 0$ for all $k$.

To describe Schoenberg's procedure, first organize the unknown points into a $n \times p$ matrix $X$ and consider the matrix of dot products $S = XX^T$, that is, $S_{ij} = x_i x_j^T$. Then the spectral theorem for symmetric matrices yields $S = U\Lambda U^T$ for orthogonal $U$ and diagonal $\Lambda$. Thus, a set of $n$ vectors which yield $S$ is given by $\tilde{X} = U\Lambda^{1/2}$. Of course, we can only retrieve $X$ up to an orthonormal transformation. This reduces the problem to finding the dot product matrix $S$ from the interpoint distances. For this, observe

$$d_{i,j}^2 = (x_i - x_j)(x_i - x_j)^T = x_i x_i^T + x_j x_j^T - 2x_i x_j^T$$

or

$$(1.2) \qquad D_2 = s\mathbf{1}^T + \mathbf{1}s^T - 2S,$$

where $D_2$ is the $n \times n$ matrix of squared distances, $s$ is the $n \times 1$ vector of the diagonal entries of $S$, and $\mathbf{1}$ is the $n \times 1$ vector of ones. The matrix $S$ can be obtained by *double centering* $D_2$:

$$(1.3) \qquad S = -\tfrac{1}{2}HD_2H, \qquad H = I - \frac{1}{n}\mathbf{1}\mathbf{1}^T.$$

To see this, first note that, for any matrix $A$, $HAH$ centers the rows and columns to have mean 0. Consequently, $Hs\mathbf{1}^T H = H\mathbf{1}s^T H = 0$ since the rows of $s\mathbf{1}^T$ and the columns of $\mathbf{1}s^T$ are constant. Pre- and post-multiplying (1.2) by $H$, we have

$$HD_2H = -2HSH.$$

Since the $x$'s were chosen as centered, $X^T\mathbf{1} = 0$, the row sums of $S$ satisfy

$$\sum_j x_i x_j^T = x_i \left(\sum_j x_j\right)^T = 0$$

and so $S = -\tfrac{1}{2}HD_2H$ as claimed.

In summary, given an $n \times n$ matrix of interpoint distances, one can solve for points achieving these distances by the following:

1. Double centering the interpoint distance squared matrix: $S = -\tfrac{1}{2}HD_2H$.
2. Diagonalizing $S$: $S = U\Lambda U^T$.
3. Extracting $\tilde{X}$: $\tilde{X} = U\Lambda^{1/2}$.

*Approximate distance matrices*: The analysis above assumes that one starts with points $x_1, x_2, \ldots, x_n$ in a $p$-dimensional Euclidean space. We may want to find an embedding $x_i \Longrightarrow y_i$ in a space of dimension $k < p$ that preserves the interpoint distances as closely as possible. Assume that $S = U\Lambda U^T$ is such that the diagonal entries of $\Lambda$ are decreasing. Set $Y_k$ to be the matrix obtained by taking the first $k$ columns of the $U$ and scaling them so that their squared norms are equal to the eigenvalues $\Lambda_k$. In particular, this provides the first $k$ columns of $X$ above and solves the minimization problem

$$(1.4) \qquad \min_{y_i \in \mathbb{R}^k} \sum_{i,j} (\|x_i - x_j\|_2^2 - \|y_i - y_j\|_2^2).$$

Young and Householder (1938) showed that this minimization can be realized as an eigenvalue problem; see the proof in this context in Mardia, Kent and Bibby (1979), page 407. In applications, an observed matrix $D$ is often not based on Euclidean distances (but may represent "dissimilarities," or just the difference of ranks). Then, the MDS solution is a heuristic for finding points in a Euclidean space whose interpoint distances approximate the orders of the dissimilarities $D$. This is called nonmetric MDS [Shepard (1962)].

*Kernel methods*: MDS converts similarities into inner products, whereas modern kernel methods [Schölkopf, Smola and Muller (1998)] start with a given matrix of inner products. Williams (2000) pointed out that Kernel PCA [Schölkopf, Smola and Muller (1998)] is equivalent to metric MDS in feature space when the kernel function is chosen isotropic, that is, the kernel $K(x, y)$ only depends on the norm $\|x - y\|$. The kernels we focus on in this paper have that property. We will show a decomposition of the horseshoe phenomenon for one particular isotropic kernel, the one defined by the kernel function $k(x_i, x_j) = \exp(-\theta(x_i - x_j)'(x_i - x_j))$.

*Relating the eigenfunctions of $S$ to those of $D_2$*: In practice, it is easier to think about the eigenfunctions of the squared distances matrix $D_2$ rather than the recentered matrix $S = -\frac{1}{2} H D_2 H$.

Observe that if $v$ is any vector such that $\mathbf{1}^T v = 0$ (i.e., the entries of $v$ sum to 0), then

$$Hv = \left(I - \frac{1}{n} \mathbf{1}\mathbf{1}^T\right)v = v.$$

Now, suppose $w$ is an eigenfunction of $D_2$ with eigenvalue $\lambda$, and let

$$\bar{w} = \left(\frac{1}{n}\sum_{i=1}^{n} w_i\right)\mathbf{1}$$

be the constant vector whose entries are the mean of $w$. Then $\mathbf{1}^T(w - \bar{w}) = 0$ and

$$S(w - \bar{w}) = -\frac{1}{2} H D_2 H(w - \bar{w})$$
$$= -\frac{1}{2} H D_2 (w - \bar{w})$$

$$= -\frac{1}{2}H(\lambda w - \lambda \bar{w} + \lambda \bar{w} - D_2 \bar{w})$$

$$= -\frac{\lambda}{2}(w - \bar{w}) + \frac{1}{2}\left(\frac{1}{n}\sum_{i=1}^{n}w_i\right)\begin{bmatrix} r_1 - \bar{r} \\ \vdots \\ r_n - \bar{r} \end{bmatrix},$$

where $r_i = \sum_{j=1}^{n}(D_2)_{ij}$ and $\bar{r} = (1/n)\sum_{i=1}^{n}r_i$. In short, if $w$ is an eigenfunction of $D_2$ and $\bar{w} = 0$, then $w$ is also an eigenfunction of $S$. By continuity, if $\bar{w} \approx 0$ or $r_i \approx \bar{r}$, then $w - \bar{w}$ is an *approximate* eigenfunction of $S$. In our setting, it turns out that the matrix $D_2$ has approximately constant row sums (so $r_i \approx \bar{r}$), and its eigenfunctions satisfy $\bar{w} \approx 0$ (in fact, some satisfy $\bar{w} = 0$). Consequently, the eigenfunctions of the centered and uncentered matrix are approximately the same in our case.

**2. A model for the data.** We begin with a brief review of models for this type of data. In spatial models of roll call voting, legislators and policies are represented by points in a low-dimensional Euclidean space with votes decided by maximizing a deterministic or stochastic utility function (each legislator choosing the policy maximizing their utility). For a precise description of these techniques, see de Leeuw (2005), where he treats the particular case of roll call data such as ours.

Since Coombs (1964), it has been understood that there is usually a natural left-right (i.e., unidimensional) model for political data. Recent comparisons [Burden, Caldeira and Groseclose (2000)] between the available left-right indices have shown that there is little difference, and that indices based on multidimensional scaling [Heckman and Snyder (1997)] perform well. Further, Heckman and Snyder (1997) conclude "standard roll call measures are good proxies of personal ideology and are still among the best measures available."

In empirical work it is often convenient to specify a parametric family of utility functions. In that context, the central problem is then to estimate those parameters and to find "ideal points" for both the legislators and the policies. A robust Bayesian procedure for parameter estimation in spatial models of roll call data was introduced in Clinton, Jackman and Rivers (2004), and provides a statistical framework for testing models of legislative behavior.

Our cut-point model is a bit different and is explained next. Although the empirical distance (1.1) is arguably a natural one to use on our data, we further motivate this choice by considering a theoretical model in which legislators lie on a regular grid in a unidimensional policy space. In this idealized model it is natural to identify legislators $l_i$ $1 \leq i \leq n$ with points in the interval $I = [0, 1]$ in correspondence with their political ideologies. We define the distance between legislators to be

$$d(l_i, l_j) = |l_i - l_j|.$$

This assumption that legislators can be isometrically mapped into an interval is key to our analysis. In the "cut-point model" for voting, each bill $1 \leq k \leq m$ on which the legislators vote is represented as a pair

$$(C_k, P_k) \in [0, 1] \times \{0, 1\}.$$

We can think of $P_k$ as indicating whether the bill is liberal ($P_k = 0$) or conservative ($P_k = 1$), and we can take $C_k$ to be the cut-point between legislators that vote "yea" or "nay." Let $V_{ik} \in \{1/2, -1/2\}$ indicate how legislator $l_i$ votes on bill $k$. Then, in this model,

$$V_{ik} = \begin{cases} 1/2 - P_k, & l_i \leq C_k, \\ P_k - 1/2, & l_i > C_k. \end{cases}$$

As described, the model has $n + 2m$ parameters, one for each legislator and two for each bill. These parameters are not identifiable without further restrictions. Adding $\varepsilon$ to $l_i$ and $C_k$ results in the same votes. Below we fix this problem by specifying values for $l_i$ and a distribution on $\{C_k\}$.

We reduce the number of parameters by assuming that the cut-points are independent random variables uniform on $I$. Then,

$$(2.1) \qquad \mathbb{P}(V_{ik} \neq V_{jk}) = d(l_i, l_j),$$

since legislators $l_i$ and $l_j$ take opposites sides on a given bill if and only if the cut-point $C_k$ divides them. Observe that the parameters $P_k$ do not affect the probability above.

The empirical distance (1.1) between legislators $l_i$ and $l_j$ generalizes to

$$\hat{d}_m(l_i, l_j) = \frac{1}{m} \sum_{k=1}^{m} |V_{ik} - V_{jk}| = \frac{1}{m} \sum_{k=1}^{m} \mathbb{1}_{V_{ik} \neq V_{jk}}.$$

By (2.1), we can estimate the latent distance $d$ between legislators by the empirical distance $\hat{d}$ which is computable from the voting record. In particular,

$$\lim_{m \to \infty} \hat{d}_m(l_i, l_j) = d(l_i, l_j) \qquad \text{a.s.},$$

since we assumed the cut-points are independent. More precisely, we have the following result:

LEMMA 2.1. *For $m \geq \log(n/\sqrt{\varepsilon})/\varepsilon^2$,*

$$\mathbb{P}(|\hat{d}_m(l_i, l_j) - d(l_i, l_j)| \leq \varepsilon \ \forall 1 \leq i, j \leq n) \geq 1 - \varepsilon.$$

PROOF. By the Hoeffding inequality, for fixed $l_i$ and $l_j$,

$$\mathbb{P}(|\hat{d}_m(l_i, l_j) - d(l_i, l_j)| > \varepsilon) \leq 2e^{-2m\varepsilon^2}.$$

Consequently,

$$\mathbb{P}\left(\bigcup_{1 \leq i < j \leq n} |\hat{d}_m(l_i, l_j) - d(l_i, l_j)| > \varepsilon\right) \leq \sum_{1 \leq i < j \leq n} \mathbb{P}(|\hat{d}_m(l_i, l_j) - d(l_i, l_j)| > \varepsilon)$$
$$\leq \binom{n}{2} 2e^{-2m\varepsilon^2}$$
$$\leq \varepsilon$$

for $m \geq \log(n/\sqrt{\varepsilon})/\varepsilon^2$, and the result follows. $\square$

We identify legislators with points in the interval $I = [0, 1]$ and define the distances between them to be $d(l_i, l_j) = |l_i - l_j|$. This general description seems to be reasonable not only for applications in political science, but also in a number of other settings. The points and the exact distance $d$ are usually unknown, however, one can often estimate $d$ from the data. For our work, we assume that one has access to an empirical distance that is *locally* accurate, that is, we assume one can estimate the distance between nearby points.

To complete the description of the model, something must be said about the hypothetical legislator points $l_i$. In Section 3 we specify these so that $d(l_i, l_j) = |i/n - j/n|$. Because of the uniformity assumption on the bill parameters and Lemma 2.1, aspects of the combination of assumptions can be empirically tested. A series of comparisons between model and data (along with scientific conclusions) are given in Section 4. These show rough but good accord; see, in particular, the comparison between Figures 3, 6, 7 and Figure 9 and the accompanying commentary.

Our model is a simple, natural set of assumptions which lead to a useful analysis of these data. The assumptions of uniform distribution of bills implies identifiability of distances between legislators. Equal spacing is the mathematically simplest assumption matching the observed distances. In informal work we have tried varying these assumptions but did not find these variations led to a better understanding of the data.

## 3. Analysis of the model.

### 3.1. *Eigenfunctions and horseshoes.* In this section we analyze multidimensional scaling applied to metric models satisfying

$$d(x_i, x_j) = |i/n - j/n|.$$

This corresponds to the case in which legislators are uniformly spaced in $I$: $l_i = i/n$. Now, if all the interpoint distances were known precisely, classical scaling would reconstruct the points exactly (up to a reversal of direction). In applications, it is often not possible to have globally accurate information. Rather, one can only reasonably approximate the interpoint distances for nearby points. To reflect this limited knowledge, we work with the dissimilarity

$$P(i,j) = 1 - \exp(-d(x_i, x_j)).$$

As a matrix,

$$P = \begin{pmatrix} 0 & 1 - e^{-1/n} & \cdots & 1 - e^{-(n-1)/n} \\ 1 - e^{-1/n} & 0 & \ddots & \vdots \\ \vdots & \ddots & \ddots & 1 - e^{-1/n} \\ 1 - e^{-(n-1)/n} & \cdots & 1 - e^{-1/n} & 0 \end{pmatrix}.$$

We are interested in finding eigenfunctions for the doubly centered matrix

$$S = -\tfrac{1}{2} HPH = -\tfrac{1}{2}(P - JP - PJ + JPJ),$$

where $J = (1/n)\mathbf{1}\mathbf{1}^T$. To prove limiting results, we work with the scaled matrices $S_n = (1/n)S$. Approximate eigenfunctions for $S_n$ are found by considering a limit $K$ of the matrices $S_n$, and then solving the corresponding integral equation

$$\int_0^1 K(x,y) f(y)\, dy = \lambda f(x).$$

Standard matrix perturbation theory is then applied to recover approximate eigenfunctions for the original, discrete matrix.

When we continuize the scaled matrices $S_n$, we get the kernel defined for $(x,y) \in [0,1] \times [0,1]$

$$K(x,y) = \tfrac{1}{2}\left( e^{-|x-y|} - \int_0^1 e^{-|x-y|}\, dx - \int_0^1 e^{-|x-y|}\, dy + \int_0^1 \int_0^1 e^{-|x-y|}\, dx\, dy \right)$$

$$= \tfrac{1}{2}(e^{-|x-y|} + e^{-y} + e^{-(1-y)} + e^{-x} + e^{-(1-x)}) + e^{-1} - 2.$$

Recognizing this as a kernel similar to those in Fredholm equations of the second type suggests that there are trigonometric solutions, as we show in Theorem A.2 in the Appendix. The eigenfunctions we derive are in agreement with those arising from the voting data, lending considerable insight into our data analysis problem and, more importantly, the horseshoe phenomenon. The sequence of explicit diagonalizations and approximations developed in the Appendix leads to the main results of this section giving closed form approximations for the eigenvectors (Theorem 3.1) and eigenvalues (Theorem 3.2), the proofs of these are also in the Appendix.

THEOREM 3.1. *Consider the centered and scaled proximity matrix defined by*

$$S_n(x_i, x_j) = \frac{1}{2n}(e^{-|i-j|/n} + e^{-i/n} + e^{-(1-i/n)} + e^{-j/n} + e^{-(1-j/n)} + 2e^{-1} - 4)$$

*for* $1 \le i, j \le n$.

1. *Set* $f_{n,a}(x_i) = \cos(a(i/n - 1/2)) - (2/a)\sin(a/2)$, *where* $a$ *is a positive solution to* $\tan(a/2) = a/(2 + 3a^2)$. *Then*

$$S_n f_{n,a}(x_i) = \frac{1}{1+a^2} f_{n,a}(x_i) + R_{f,n}, \qquad \text{where } |R_{f,n}| \le \frac{a+4}{2n}.$$

2. *Set* $g_{n,a}(x_i) = \sin(a(i/n - 1/2))$, *where* $a$ *is a positive solution to* $a\cot(a/2) = -1$. *Then*

$$S_n g_{n,a}(x_i) = \frac{1}{1+a^2} g_{n,a}(x_i) + R_{g,n}, \qquad \text{where } |R_{g,n}| \le \frac{a+2}{2n}.$$

*That is,* $f_{n,a}$ *and* $g_{n,a}$ *are approximate eigenfunctions of* $S_n$.

THEOREM 3.2. *Consider the setting of Theorem 3.1 and let* $\lambda_1, \ldots, \lambda_n$ *be the eigenvalues of* $S_n$.

1. *For positive solutions to* $\tan(a/2) = a/(2 + 3a^2)$,

$$\min_{1 \le i \le n} \left| \lambda_i - \frac{1}{1+a^2} \right| \le \frac{a+4}{\sqrt{n}}.$$

2. *For positive solutions to* $a\cot(a/2) = -1$,

$$\min_{1 \le i \le n} \left| \lambda_i - \frac{1}{1+a^2} \right| \le \frac{a+2}{\sqrt{n}}.$$

In the Appendix we prove an uncentered version of this theorem (Theorem A.3) that is used in the case of uncentered matrices which we will need for the double horseshoe case of the next section.

In the results above, we transformed distances into dissimilarities via the exponential transformation $P(i, j) = 1 - \exp(-d(x_i, x_j))$. If we worked with the distances directly, so that the dissimilarity matrix is given by $P(i, j) = |l_i - l_j|$, then much of what we develop here stays true. In particular, the operators are explicitly diagonalizable with similar eigenfunctions. This has been independently studied by physicists in what they call the *crystal configuration* of a one-dimensional Anderson model, with spectral decomposition analyzed in Bogomolny, Bohigas and Schmit (2003).

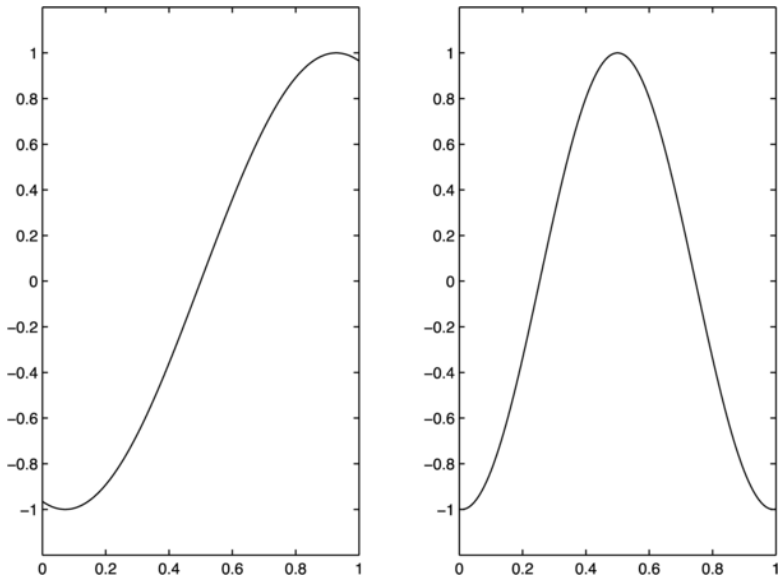

Fig. 2. *Approximate eigenfunctions $f_1$ and $f_2$.*

3.1.1. *Horseshoes and twin horseshoes.* The 2-dimensional MDS mapping is built out of the first and second eigenfunctions of the centered proximity matrix. As shown above, we have the following approximate eigenfunctions:

- $f_1(x_i) = f_{n,a_1}(x_i) = \sin(3.67(i/n - 1/2))$ with eigenvalue $\lambda_1 \approx 0.07$,
- $f_2(x_i) = f_{n,a_2}(x_i) = \cos(6.39(i/n - 1/2))$ with eigenvalue $\lambda_2 \approx 0.02$,

where the eigenvalues are for the scaled matrix. Figure 2 shows a graph of these eigenfunctions. Moreover, Figure 3 shows the horseshoe that results from plotting $\Lambda : x_i \mapsto (\sqrt{\lambda_1} f_1(x_i), \sqrt{\lambda_2} f_2(x_i))$. From $\Lambda$ it is possible to deduce the relative order of the Representatives in the interval $I$. Since $-f_1$ is also an eigenfunction, it is not in general possible to determine the absolute order knowing only that $\Lambda$ comes from the eigenfunctions. However, as can be seen in Figure 3, the relationship between the two eigenfunctions is a curve for which we have the parametrization given above, but which cannot be written in functional form, in particular, the second eigenvector is not a quadratic function of the first as is sometimes claimed.

With the voting data, we see not one, but two horseshoes. To see how this can happen, consider the two population state space $\mathcal{X} = \{x_1, \ldots, x_n, y_1, \ldots, y_n\}$ with proximity $d(x_i, x_j) = 1 - e^{-|i/n - j/n|}$, $d(y_i, y_j) = 1 - e^{-|i/n - j/n|}$ and $d(x_i, y_j) = 1$. This leads to the partitioned proximity matrix

$$\tilde{P}_{2n} = \left[ \begin{array}{c|c} P_n & 1 \\ \hline 1 & P_n \end{array} \right],$$

where $P_n(i,j) = 1 - e^{-|i/n - j/n|}$.

COROLLARY 3.1. *From Theorem A.3 we have the following approximate eigenfunctions and eigenvalues for* $-(1/2n)\tilde{P}_{2n}$:

- $f_1(i) = \cos(a_1(i/n - 1/2))$, *for* $1 \leq i \leq n$ $f_1(j) = -\cos(a_1((j-n)/n - 1/2))$ *for* $(n+1) \leq j \leq 2n$, *where* $a_1 \approx 1.3$ *and* $\lambda_1 \approx 0.37$.
- $f_2(i) = \sin(a_2(i/n - 1/2))$, *for* $1 \leq i \leq n$ $f_2(j) = 0$ *for* $(n+1) \leq j \leq 2n$, *where* $a_2 \approx 3.67$ *and* $\lambda_2 \approx 0.069$.
- $f_3(i) = 0$, *for* $1 \leq i \leq n$, $f_3(j) = \sin(a_2((j-n)/n - 1/2))$ *for* $(n+1) \leq j \leq 2n$, *where* $a_2 \approx 3.67$ *and* $\lambda_3 \approx 0.069$.

PROOF.

$$-\frac{1}{2n}\tilde{P}_{2n} = \left[\begin{array}{c|c} A_n & 0 \\ \hline 0 & A_n \end{array}\right] - \frac{1}{2n}\mathbf{1}\mathbf{1}^T,$$

where $A_n(i,j) = (1/2n)e^{-|i/n - j/n|}$. If $u$ is an eigenvector of $A_n$, then the vector $(u, -u)$ of length $2n$ is an eigenvector of $-\frac{1}{2n}\tilde{P}_{2n}$ since

$$\left(\left[\begin{array}{c|c} A_n & 0 \\ \hline 0 & A_n \end{array}\right] - \frac{1}{2n}\mathbf{1}\mathbf{1}^T\right)\begin{pmatrix} u \\ -u \end{pmatrix} = \lambda_1\begin{pmatrix} u \\ -u \end{pmatrix} + 0.$$

If we additionally have that $\mathbf{1}^T u = 0$, then, similarly, $(u, \vec{0})$ and $(\vec{0}, u)$ are also eigenfunctions of $-\frac{1}{2n}\tilde{P}_{2n}$. $\quad\square$

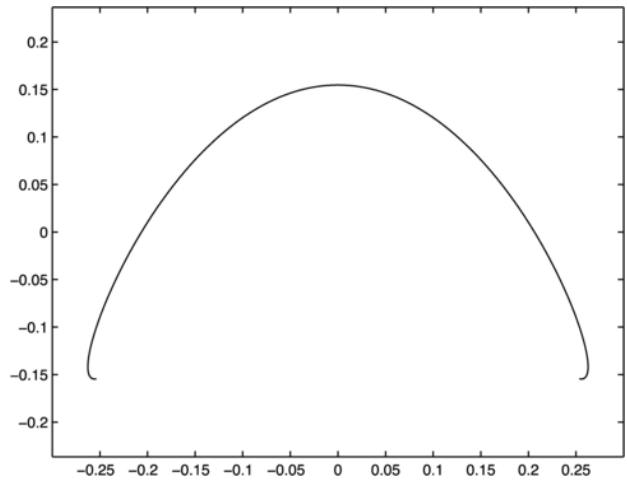

FIG. 3. *A horseshoe that results from plotting* $\Lambda : x_i \mapsto (\sqrt{\lambda_1}f_1(x_i), \sqrt{\lambda_2}f_2(x_i))$.

Since the functions $f_1$, $f_2$ and $f_3$ of Corollary 3.1 are all orthogonal to constant functions, by the discussion in Section 1.2 they are also approximate eigenfunctions for the centered, scaled matrix $(-1/2n)H\tilde{P}_{2n}H$. These functions are graphed in Figure 4, and the twin horseshoes that result from the 3-dimensional mapping $\Lambda : z \mapsto (\sqrt{\lambda_1}f_1(z), \sqrt{\lambda_2}f_2(z), \sqrt{\lambda_3}f_3(z))$ are shown in Figure 5. The first eigenvector provides the separation into two groups, this is a well known method for separating clusters known today as spectral clustering [Shi and Malik (2000)]. For a nice survey and consistency results see von Luxburg, Belkin and Bousquet (2008).

REMARK. The matrices $A_n$ and $\tilde{P}_{2n}$ above are centrosymmetric [Weaver (1985)], that is, symmetrical around the center of the matrix. Formally, if $K$ is the matrix with 1's in the counter (or secondary) diagonal,

$$K = \begin{pmatrix} 0 & 0 & \dots & 0 & 1 \\ 0 & 0 & \dots & 1 & 0 \\ \vdots & & \vdots & & \\ 0 & 1 & \dots & 0 & 0 \\ 1 & 0 & \dots & 0 & 0 \end{pmatrix},$$

then a matrix $B$ is centrosymmetric iff $BK = KB$. A very useful review by Weaver (1985) quotes I. J. Good (1970) on the connection between centrosymmetric matrices and kernels of integral equations: *"Toeplitz matrices (which are examples of matrices which are both symmetric and centrosymmetric) arise as discrete approximations to kernels $k(x,t)$ of integral equations when these kernels are functions of $|x - t|$."* (Today we would call

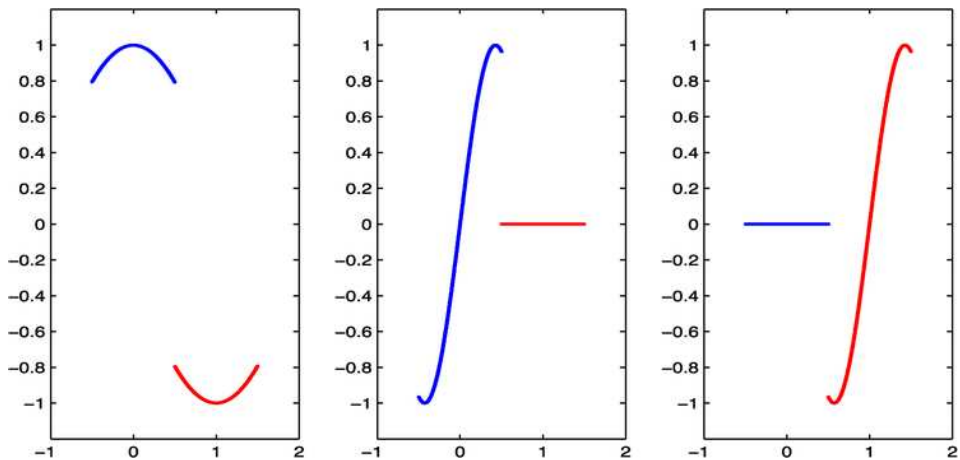

FIG. 4. *Approximate eigenfunctions $f_1$, $f_2$ and $f_3$ for the centered proximity matrix arising from the two population model.*

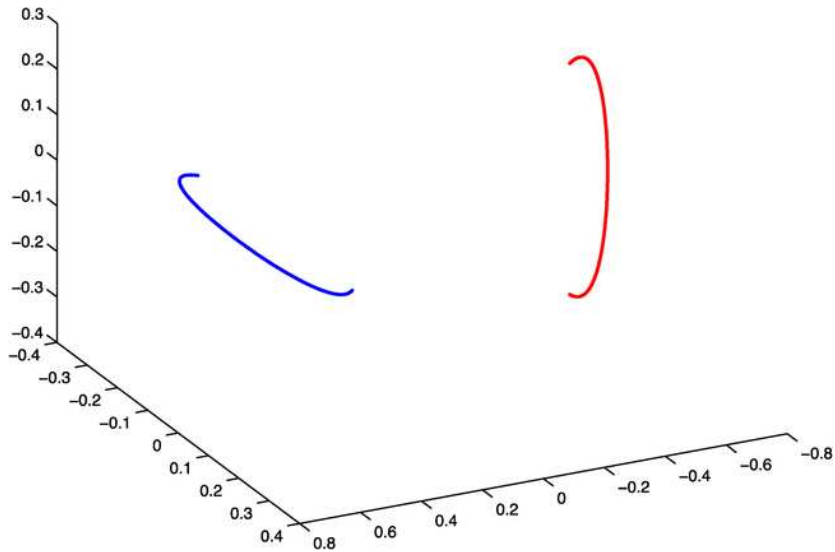

Fig. 5. *Twin horseshoes in the two population model that result from plotting* $\Lambda\colon z \mapsto (\sqrt{\lambda_1} f_1(z), \sqrt{\lambda_2} f_2(z), \sqrt{\lambda_3} f_3(z))$.

*these isotropic kernels.)* *"Similarly if a kernel is an even function of its vector argument (x, t), that is, if $k(x,t) = k(-x,-t)$, then it can be discretely approximated by a centrosymmetric matrix."*

Centrosymmetric matrices have very neat eigenvector formulas [Cantoni and Butler (1976)]. In particular, if the order of the matrix, $n$, is even, then the first eigenvector is skew symmetric and thus of the form $(u_1, -u_1)$ and orthogonal to the constant vector. This explains the miracle that seems to occur in the simplification of the eigenvectors in the above formulae.

**4. Connecting the model to the data.** When we apply MDS to the voting data, the first three eigenvalues are as follows:

- 0.13192,
- 0.00764,
- 0.00634.

Observe that as our two population model suggests, the second and third eigenvalues are about equal and significantly smaller than the first.

Figure 6 shows the first, second and third eigenfunctions $f_1$, $f_2$ and $f_3$ from the voting data. The 3-dimensional MDS plot in Figure 1(a) is the graph of $\Lambda\colon x_i \mapsto (\sqrt{\lambda_1} f_1(x_i), \sqrt{\lambda_2} f_2(x_i), \sqrt{\lambda_3} f_3(x_i))$. Since legislators are not a priori ordered, the eigenfunctions are difficult to interpret. However, our model suggests the following ordering: Split the legislators into two groups $G_1$ and $G_2$ based on the sign of $f_1(x_i)$; then the norm of $f_2$ is larger on one group, say, $G_1$, so we sort $G_1$ based on increasing values of $f_2$, and similarly, sort $G_2$ via $f_3$. Figure 7 shows the same data as does Figure 6, but with this judicious ordering of the legislators. Figure 8 shows the ordered eigenfunctions obtained from MDS applied to the 2004 roll call data. The results appear to be in agreement with the theoretically derived functions in Figure 4. This agreement gives one validation of the modeling assumptions in Section 2.

The theoretical second and third eigenfunctions are part of a two-dimensional eigenspace. In the voting data it is reasonable to assume that noise eliminates symmetry and collapses the eigenspaces down to one dimension. Nonetheless, we would guess that the second and third eigenfunctions in the voting data are in the two-dimensional predicted eigenspace, as is seen to be the case in Figures 7 and 8.

Our analysis in Section 3 suggests that if legislators are in fact isometrically embedded in the interval $I$ (relative to the roll call distance), then their MDS derived rank will be consistent with the order of legislators in the interval. This appears to be the case in the data, as seen in Figure 9, which shows a graph of $\hat{d}(l_i, \cdot)$ for selected legislators $l_i$. For example, as we would predict, $\hat{d}(l_1, \cdot)$ is an increasing function and $\hat{d}(l_n, \cdot)$ is decreasing. Moreover, the data seem to be in rough agreement with the metric assumption of our two population model, namely, that the two groups are well separated and that the within group distance is given by $d(l_i, l_j) = |i/n - j/n|$. This agreement is another validation of the modeling assumptions in Section 2.

Our voting model suggests that the MDS ordering of legislators should correspond to political ideology. To test this, we compared the MDS re-sults to the assessment of legislators by Americans for Democratic Action [Americans for Democratic Action (2005)]. Each year ADA selects 20 votes it considers the most important during that session, for example, the Patriot Act reauthorization. Legislators are assigned a Liberal Quotient: the percentage of those 20 votes on which the Representative voted in accor-dance with what ADA considered to be the liberal position. For example, a representative who voted the liberal position on all 20 votes would receive an LQ of 100%. Figure 10 below shows a plot of LQ vs. MDS rank.

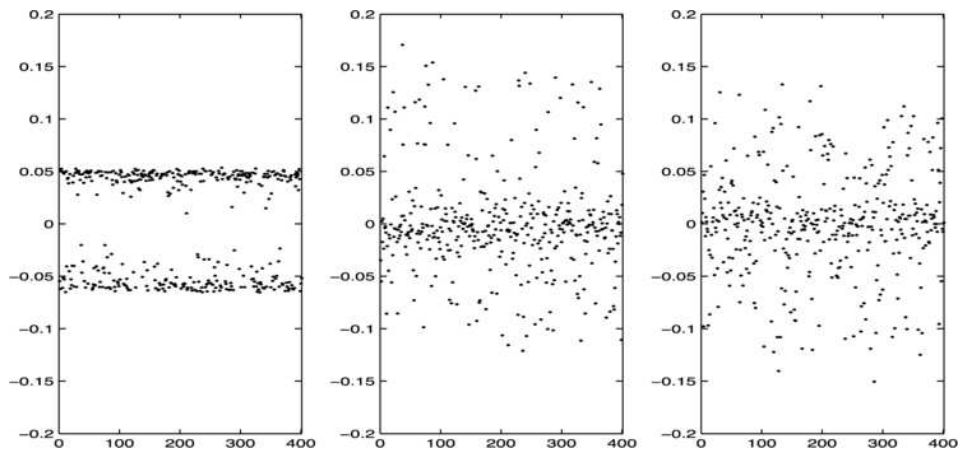

FIG. 6.    *The first, second and third eigenfunctions output from MDS applied to the 2005 U.S. House of Representatives roll call votes.*

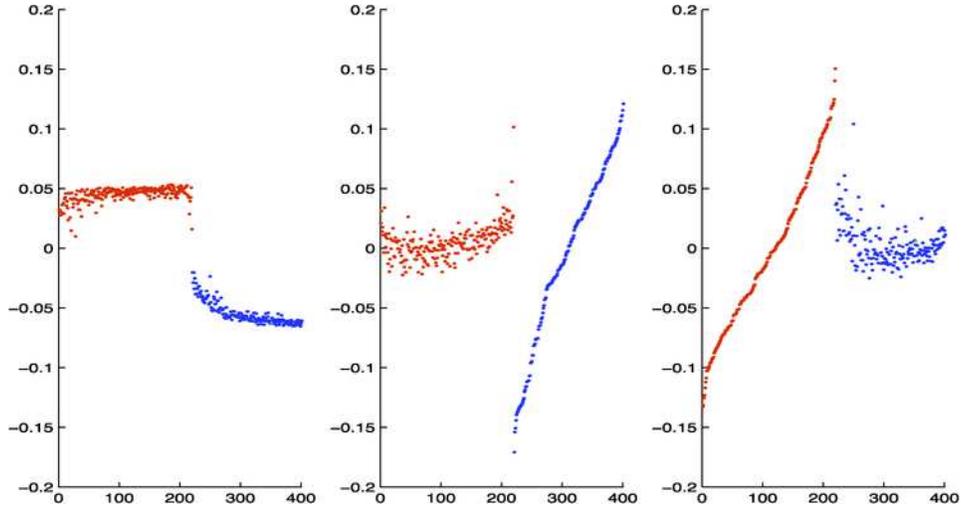

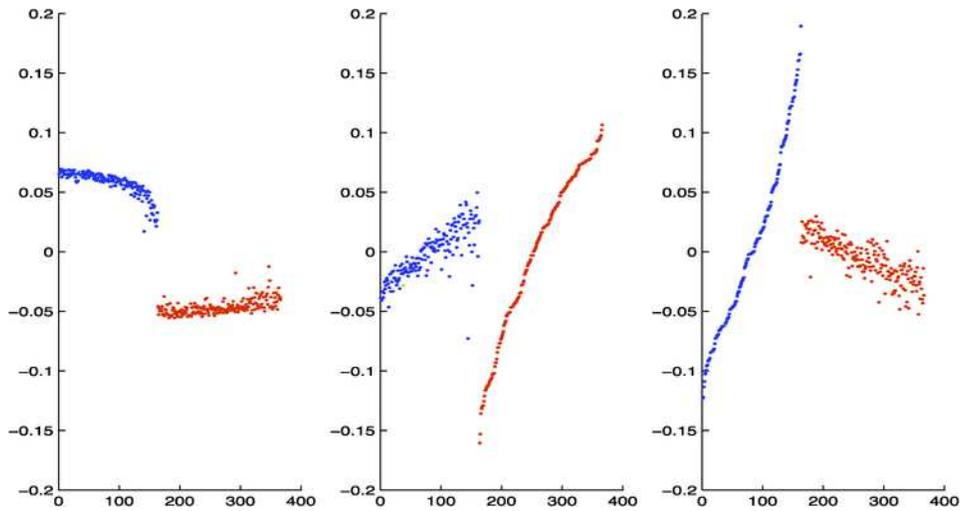

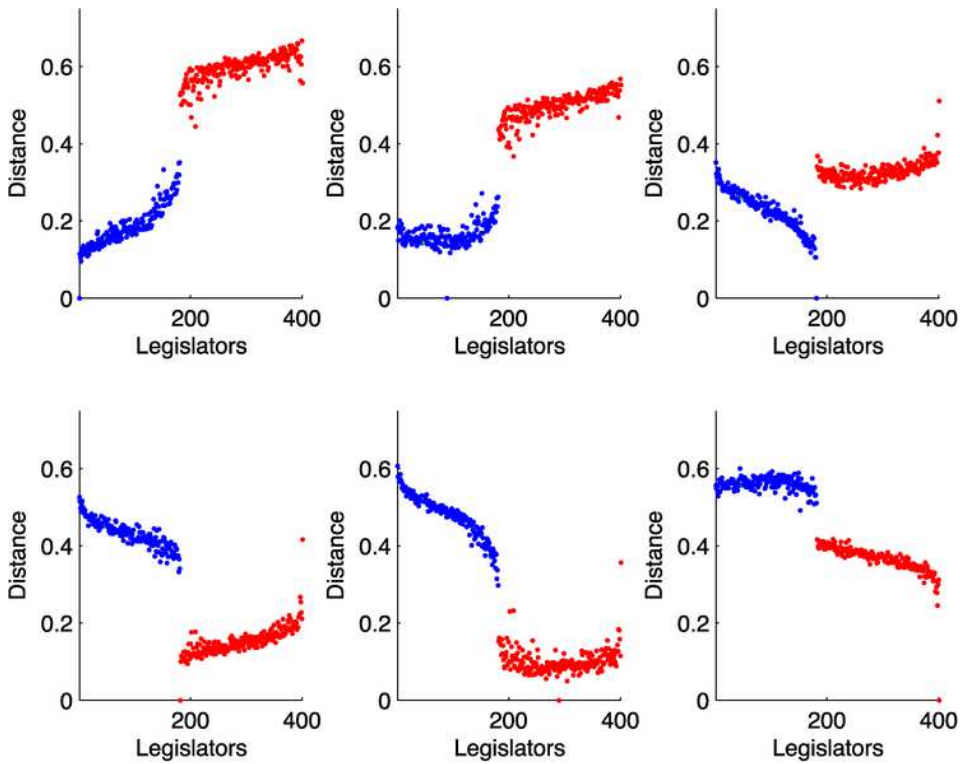

FIG. 9.  *The empirical roll call derived distance function $\hat{d}(l_i, \cdot)$ for selected legislators $l_i = 1, 90, 181, 182, 290, 401$. The x-axis orders legislators according to their MDS rank.*

For the most part, the two measures are consistent. However, MDS separates two groups of relatively liberal Republicans. To see why this is the case, consider the two legislators Mary Bono (R-CA) with MDS rank 248 and Gil Gutknecht (R-MN) with rank 373. Both Representatives received an ADA rating of 15%, yet had considerably different voting records. On the 20 ADA bills, both Bono and Gutknecht supported the liberal position 3 times—but never simultaneously. Consequently, the empirical roll call distance between them is relatively large considering that they are both Republicans. Since MDS attempts to preserve local distances, Bono and Gutknecht are consequently separated by the algorithm. In this case, distance is directly related to the propensity of legislators to vote the same on any given bill. Figure 10 results because this notion of proximity, although related, does not correspond directly to political ideology. The MDS and ADA rankings complement one another in the sense that together they facilitate identification of two distinct, yet relatively liberal groups of Republicans. That is, although these two groups are relatively liberal, they do not share the same political positions.

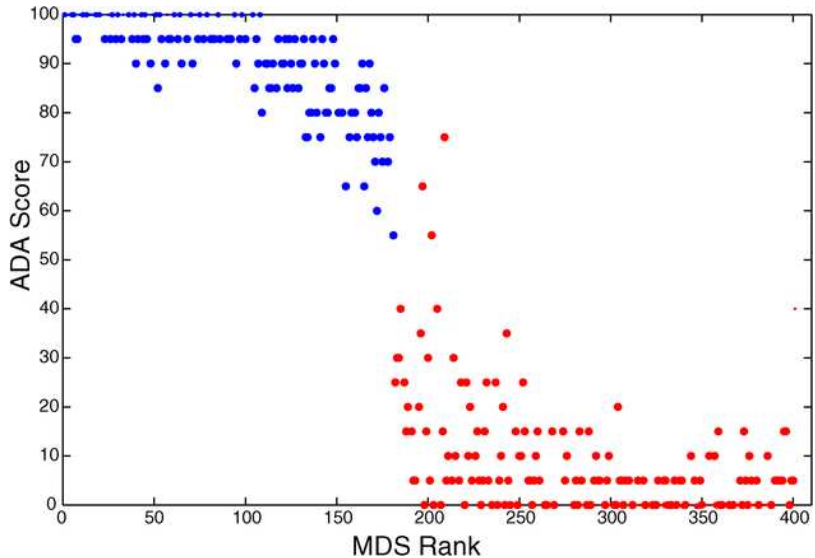

Fig. 10. *Comparison of the MDS derived rank for Representatives with the Liberal Quotient as defined by Americans for Democratic Action.*

Like ADA, the National Journal ranks Representatives each year based on their voting record. In 2005, The Journal chose 41 votes on economic issues, 42 on social issues and 24 dealing with foreign policy. Based on these 107 votes, legislators were assigned a rating between 0 and 100—lower numbers indicate a more liberal political ideology. Figure 11 is a plot of the National Journal vs. MDS rankings, and shows results similar to the ADA comparison. As in the ADA case, we see that relatively liberal Republicans receive quite different MDS ranks. Interestingly, this phenomenon does not appear for Democrats under either the ADA or the National Journal ranking system.

**Summary.** Our work began with an empirical finding: multidimensional scaling applied to voting data from the US house of representatives shows a clean double horseshoe pattern (Figure 1). These patterns happen often enough in data reduction techniques that it is natural to seek a theoretical understanding. Our main results give a limiting closed form explanation for data matrices that are double-centered versions of

$$P(i,j) = 1 - e^{-\theta|i/n - j/n|}, \qquad 1 \le i, j \le n.$$

We further show how voting data arising from a cut-point model developed in Section 3 gives rise to a model of this form.

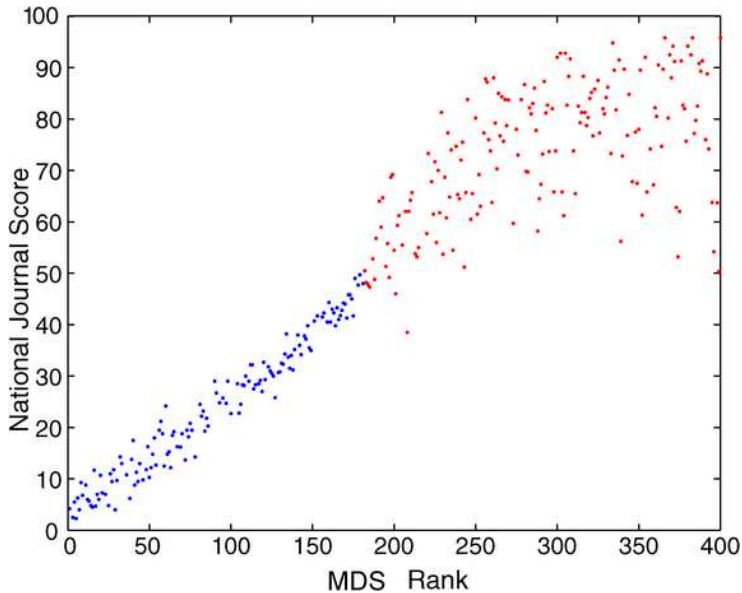

Fig. 11.  *Comparison of the eigendecomposition derived rank for Representatives with the National Journal's liberal score.*

In a followup to this paper, de Leeuw (2007) has shown that some of our results can be derived directly without passing to a continuous kernel. A useful byproduct of his results and conversations with colleagues and students is this: the matrix $P_{i,j}$ above is totally positive. Standard theory shows that the first eigenvector can be taken increasing and the second as unimodal. Plotting these eigenvectors versus each other will always result in a horseshoe shape. Perhaps this explains the ubiquity of horseshoes.

## APPENDIX: THEOREMS AND PROOFS FOR SECTION 3

We state first a classical perturbation result that relates two different notions of an *approximate eigenfunction*. A proof is included here to aid the reader. For more refined estimates, see Parlett (1980), Chapter 4, page 69.

Two lemmas provide trigonometric identities that are useful for finding the eigenfunctions for the continuous kernel. Theorem A.2 states specific solutions to this integral equation. We then provide a proof for Theorem 3.1. The version of this theorem for uncentered matrices (Theorem A.3) follows and is used in the two horseshoe case.

THEOREM A.1.   *Consider an $n \times n$ symmetric matrix $A$ with eigenvalues $\lambda_1 \leq \cdots \leq \lambda_n$. If for $\varepsilon > 0$*

$$\|Af - \lambda f\|_2 \leq \varepsilon$$

*for some $f, \lambda$ with $\|f\|_2 = 1$, then $A$ has an eigenvalue $\lambda_k$ such that $|\lambda_k - \lambda| \leq \varepsilon$.*

*If we further assume that*

$$s = \min_{i:\lambda_i \neq \lambda_k} |\lambda_i - \lambda_k| > \varepsilon,$$

*then $A$ has an eigenfunction $f_k$ such that $Af_k = \lambda_k f_k$ and $\|f - f_k\|_2 \leq \varepsilon/(s - \varepsilon)$.*

PROOF. First we show that $\min_i |\lambda_i - \lambda| \leq \varepsilon$. If $\min_i |\lambda_i - \lambda| = 0$, we are done; otherwise $A - \lambda I$ is invertible. Then,

$$\|f\|_2 \leq \|(A - \lambda I)^{-1}\| \cdot \|(A - \lambda)f\|_2$$
$$\leq \varepsilon \|(A - \lambda I)^{-1}\|.$$

Since the eigenvalues of $(A - \lambda I)^{-1}$ are $1/(\lambda_1 - \lambda), \ldots, 1/(\lambda_n - \lambda)$, by symmetry,

$$\|(A - \lambda I)^{-1}\| = \frac{1}{\min_i |\lambda_i - \lambda|}.$$

The result now follows since $\|f\|_2 = 1$.

Set $\lambda_k = \operatorname{argmin}|\lambda_i - \lambda|$, and consider an orthonormal basis $g_1, \ldots, g_m$ of the associated eigenspace $E_{\lambda_k}$. Define $f_k$ to be the projection of $f$ onto $E_{\lambda_k}$:

$$f_k = \langle f, g_1 \rangle g_1 + \cdots + \langle f, g_m \rangle g_m.$$

Then $f_k$ is an eigenfunction with eigenvalue $\lambda_k$. Writing $f = f_k + (f - f_k)$, we have

$$(A - \lambda I)f = (A - \lambda I)f_k + (A - \lambda I)(f - f_k)$$
$$= (\lambda_k - \lambda)f_k + (A - \lambda I)(f - f_k).$$

Since $f - f_k \in E_{\lambda_k}^{\perp}$, by symmetry, we have

$$\langle f_k, A(f - f_k) \rangle = \langle Af_k, f - f_k \rangle = \langle \lambda_k f_k, f - f_k \rangle = 0.$$

Consequently, $\langle f_k, (A - \lambda I)(f - f_k) \rangle = 0$ and by Pythagoras,

$$\|Af - \lambda f\|_2^2 = (\lambda_k - \lambda)^2 \|f_k\|_2 + \|(A - \lambda I)(f - f_k)\|_2^2.$$

In particular,

$$\varepsilon \geq \|Af - \lambda f\|_2 \geq \|(A - \lambda I)(f - f_k)\|_2.$$

For $\lambda_i \neq \lambda_k$, $|\lambda_i - \lambda| \geq s - \varepsilon$. The result now follows since for $h \in E_{\lambda_k}^{\perp}$

$$\|(A - \lambda I)h\|_2 \geq (s - \varepsilon)\|h\|_2. \qquad \square$$

REMARK A.1. The second statement of the theorem allows nonsimple eigenvalues, but requires that the eigenvalues corresponding to distinct eigenspaces be well separated.

REMARK A.2. The eigenfunction bound of the theorem is asymptotically tight in $\varepsilon$ as the following example illustrates: Consider the matrix

$$A = \begin{bmatrix} \lambda & 0 \\ 0 & \lambda + s \end{bmatrix}$$

with $s > 0$. For $\varepsilon < s$, define the function

$$f = \begin{bmatrix} \sqrt{1 - \varepsilon^2/s^2} \\ \varepsilon/s \end{bmatrix}.$$

Then $\|f\|_2 = 1$ and $\|Af - \lambda f\|_2 = \varepsilon$. The theorem guarantees that there is an eigenfunction $f_k$ with eigenvalue $\lambda_k$ such that $|\lambda - \lambda_k| \le \varepsilon$. Since the eigenvalues of $A$ are $\lambda$ and $\lambda + s$, and since $s > \varepsilon$, we must have $\lambda_k = \lambda$. Let $V_k = \{f_k : Af_k = \lambda_k f_k\} = \{ce_1 : c \in \mathbb{R}\}$, where $e_1$ is the first standard basis vector. Then

$$\min_{f_k \in V_k} \|f - f_k\|_2 = \|f - (f \cdot e_1)e_1\| = \varepsilon/s.$$

The bound of the theorem, $\varepsilon/(s - \varepsilon)$, is only slightly larger.

We establish an integral identity in order to find trigonometric solutions to $Kf = \lambda f$ where $K$ is the continuized kernel of the centered exponential proximity matrix.

LEMMA A.1. *For constants $a \in \mathbb{R}$ and $c \in [0, 1]$,*

$$\int_0^1 e^{-|x-c|} \cos[a(x - 1/2)]\,dx$$

$$= \frac{2\cos[a(c - 1/2)]}{1 + a^2} + \frac{(e^{-c} + e^{c-1})(a\sin(a/2) - \cos(a/2))}{1 + a^2}$$

*and*

$$\int_0^1 e^{-|x-c|} \sin[a(x - 1/2)]\,dx$$

$$= \frac{2\sin[a(c - 1/2)]}{1 + a^2} + \frac{(e^{-c} - e^{c-1})(a\cos(a/2) + \sin(a/2))}{1 + a^2}.$$

PROOF. The lemma follows from a straightforward integration. First split the integral into two pieces:

$$\int_0^1 e^{-|x-c|} \cos[a(x - 1/2)]\,dx$$

$$= \int_0^c e^{x-c} \cos[a(x - 1/2)]\,dx + \int_c^1 e^{c-x} \cos[a(x - 1/2)]\,dx.$$

By integration by parts applied twice,

$$\int e^{x-c} \cos[a(x-1/2)]\,dx = \frac{ae^{x-c}\sin(a(x-1/2)) + e^{x-c}\cos(a(x-1/2))}{1+a^2}$$

and

$$\int e^{c-x} \cos[a(x-1/2)]\,dx = \frac{ae^{c-x}\sin(a(x-1/2)) - e^{c-x}\cos(a(x-1/2))}{1+a^2}.$$

Evaluating these expressions at the appropriate limits of integration gives the first statement of the lemma. The computation of $\int_0^1 e^{-|x-c|}\sin[a(x-1/2)]\,dx$ is analogous, and so is omitted here. $\square$

We now derive eigenfunctions for the continuous kernel.

THEOREM A.2. *For the kernel*

$$K(x,y) = \tfrac{1}{2}(e^{-|x-y|} + e^{-y} + e^{-(1-y)} + e^{-x} + e^{-(1-x)}) + e^{-1} - 2$$

*defined on* $[0,1] \times [0,1]$, *the corresponding integral equation*

$$\int_0^1 K(x,y)f(y)\,dy = \lambda f(x)$$

*has solutions*

$$f(x) = \sin(a(x-1/2)), \qquad a\cot(a/2) = -1$$

*and*

$$f(x) = \cos(a(x-1/2)) - \frac{2}{a}\sin(a/2), \qquad \tan(a/2) = \frac{a}{2+3a^2}.$$

*In both cases,* $\lambda = 1/(1+a^2)$.

PROOF. First note that both classes of functions in the statement of the theorem satisfy $\int_0^1 f(x)\,dx = 0$. Consequently, the integral simplifies to

$$\int_0^1 K(x,y)f(y)\,dy = \tfrac{1}{2}\int_0^1 (e^{-|x-y|} + e^{-y} + e^{-(1-y)})f(y)\,dy.$$

Furthermore, since $e^{-y} + e^{-(1-y)}$ is symmetric about $1/2$ and $\sin(a(y-1/2))$ is skew-symmetric about $1/2$, Lemma A.1 shows that

$$\int_0^1 K(x,y)\sin(a(y-1/2))\,dy$$

$$= \frac{1}{2}\int_0^1 e^{-|x-y|}\sin(a(y-1/2))\,dy$$

$$= \frac{\sin[a(c-1/2)]}{1+a^2} + \frac{(e^{-c}-e^{c-1})(a\cos(a/2)+\sin(a/2))}{2(1+a^2)}.$$

This establishes the first statement of the theorem. We examine the second. Since $\int_0^1 K(x,y)\,dy = 0$,

$$\int_0^1 (e^{-|x-y|} + e^{-y} + e^{-(1-y)})\,dy = (4 - 2e^{-1} - e^{-x} - e^{-(1-x)})$$

and also, by straightforward integration by parts,

$$\int_0^1 e^{-y}\cos(a(y-1/2))\,dy = \int_0^1 e^{-(1-y)}\cos(a(y-1/2))\,dy$$

$$= \frac{a\sin(a/2)(1+e^{-1})}{1+a^2} + \frac{\cos(a/2)(1-e^{-1})}{1+a^2}.$$

Using the result of Lemma [A.1], we have

$$\frac{1}{2}\int_0^1 [e^{-|x-y|} + e^{-y} + e^{-(1-y)}]\Big[\cos(a(y-1/2)) - \frac{2}{a}\sin(a/2)\Big]\,dy$$

$$= \frac{\cos[a(x-1/2)]}{1+a^2} + \frac{(e^{-x} + e^{x-1})(a\sin(a/2) - \cos(a/2))}{2(1+a^2)}$$

$$\quad + \frac{a\sin(a/2)(1+e^{-1})}{1+a^2} + \frac{\cos(a/2)(1-e^{-1})}{1+a^2}$$

$$\quad - \frac{1}{a}\sin(a/2)(4 - 2e^{-1} - e^{-x} - e^{-(1-x)})$$

$$= \frac{\cos[a(x-1/2)]}{1+a^2} - \frac{2\sin(a/2)}{a(1+a^2)} + \frac{\phi(x)}{a(1+a^2)},$$

where

$$\phi(x) = 2\sin(a/2) + a(e^{-x} + e^{x-1})(a\sin(a/2) - \cos(a/2))/2$$

$$\quad + a^2\sin(a/2)(1+e^{-1}) + a\cos(a/2)(1-e^{-1})$$

$$\quad - (1+a^2)\sin(a/2)(4 - 2e^{-1} - e^{-x} - e^{-(1-x)}).$$

The result follows by grouping the terms of $\phi(x)$ so that we see

$$\phi(x) = [2 - 4 + 2e^{-1} + e^{-x} + e^{-(1-x)}]\sin(a/2)$$

$$\quad + [e^{-x}/2 + e^{x-1}/2 + 1 + e^{-1} - 4 + 2e^{-1} + e^{-x} + e^{-(1-x)}]a^2\sin(a/2)$$

$$\quad + [-e^{-x}/2 - e^{x-1}/2 + 1 - e^{-1}]a\cos(a/2)$$

$$= [-e^{-x}/2 - e^{x-1}/2 + 1 - e^{-1}]$$

$$\quad \times [a\cos(a/2) - 2\sin(a/2) - 3a^2\sin(a/2)]. \qquad \square$$

Theorem [A.2] states specific solutions to our integral equation. Now we show that in fact these are all the solutions with positive eigenvalues. To start, observe that for $0 \le x, y \le 1$, $e^{-1} \le e^{-|x-y|} \le 1$ and $e^{-1} + 1 \le e^{-x} + e^{-(1-x)} \le 2e^{-1/2}$. Consequently,

$$-1 < \tfrac{3}{2}e^{-1} + 1 + e^{-1} - 2 \le K(x,y) \le \tfrac{1}{2} + 2e^{-1/2} + e^{-1} - 2 < 1$$

and so $\|K\|_\infty < 1$. In particular, if $\lambda$ is an eigenvalue of $K$, then $|\lambda| < 1$. Now suppose $f$ is an eigenfunction of $K$, that is,

$$\lambda f(x) = \int_0^1 [\tfrac{1}{2}(e^{-|x-y|} + e^{-x} + e^{-(1-x)} + e^{-y} + e^{-(1-y)}) + e^{-1} - 2] f(y) \, dy.$$

Taking the derivative with respect to $x$, we see that $f$ satisfies

(A-1)     $\lambda f'(x) = \tfrac{1}{2} \int_0^1 (-e^{-|x-y|} H_y(x) - e^{-x} + e^{-(1-x)}) f(y) \, dy,$

where $H_y(x)$ is the Heaviside function, that is, $H_y(x) = 1$ for $x \ge y$ and $H_y(x) = -1$ for $x < y$. Taking the derivative again, we get

(A-2)     $\lambda f''(x) = -f(x) + \tfrac{1}{2} \int_0^1 (e^{-|x-y|} + e^{-x} + e^{-(1-x)}) f(y) \, dy.$

Now, substituting back into the integral equation, we see

$$\lambda f(x) = \lambda f''(x) + f(x) + \int_0^1 [\tfrac{1}{2}(e^{-y} + e^{-(1-y)}) + e^{-1} - 2] f(y) \, dy.$$

Taking one final derivative with respect to $x$, and setting $g(x) = f'(x)$, we see

(A-3)                                $g''(x) = \dfrac{\lambda - 1}{\lambda} g(x).$

For $0 < \lambda < 1$, all the solutions to (A-3) can be written in the form

$$g(x) = A \sin(a(x - 1/2)) + B \cos(a(x - 1/2))$$

with $\lambda = 1/(1 + a^2)$. Consequently, $f(x)$ takes the form

$$f(x) = A \sin(a(x - 1/2)) + B \cos(a(x - 1/2)) + C.$$

Note that since $\int_0^1 K(x,y) \, dy = 0$, the constant function $c(x) \equiv 1$ is an eigenfunction of $K$ with eigenvalue 0. Since $K$ is symmetric, for any eigenfunction $f$ with nonzero eigenvalue, $f$ is orthogonal to $c$ in $L^2(dx)$, that is, $\int_0^1 f(x) \, dx = 0$. In particular, for $0 < \lambda < 1$, without loss, we assume

$$f(x) = A \sin(a(x - 1/2)) + B \left[ \cos(a(x - 1/2)) - \frac{2}{a} \sin(a/2) \right].$$

We solve for $a$, $A$ and $B$. First assume $B \ne 0$, and divide $f$ through by $B$. Then $f(1/2) = 1 - (2/a) \sin(a/2)$. Since $K(x, \cdot)$ is symmetric about $1/2$ and

$\sin(a(x - 1/2))$ is skew-symmetric about $1/2$, we have

$$\lambda f(1/2) = \frac{1 - (2/a)\sin(a/2)}{1 + a^2}$$

$$= \int_0^1 \left[\frac{1}{2}(e^{|y-1/2|} + e^{-y} + e^{-(1-y)}) + e^{-1/2} + e^{-1} - 2\right] f(y)\, dy$$

$$= \frac{1}{2} \int_0^1 (e^{|y-1/2|} + e^{-y} + e^{-(1-y)}) \cos(a(y - 1/2))\, dy$$

$$+ \frac{2}{a} \sin(a/2)(e^{-1/2} + e^{-1} - 2)$$

$$= \frac{1}{1 + a^2} + \frac{e^{-1/2}(a \sin(a/2) - \cos(a/2))}{1 + a^2}$$

$$+ \frac{a \sin(a/2)(1 + e^{-1})}{1 + a^2} + \frac{\cos(a/2)(1 - e^{-1})}{1 + a^2}$$

$$+ \frac{2}{a} \sin(a/2)(e^{-1/2} + e^{-1} - 2).$$

The last equality follows from Lemma A.1. Equating the sides, $a$ satisfies

$$0 = 2\sin(a/2) + e^{-1/2}a(a\sin(a/2) - \cos(a/2)) + a^2 \sin(a/2)(1 + e^{-1})$$

$$+ a\cos(a/2)(1 - e^{-1}) + 2(1 + a^2)\sin(a/2)(e^{-1/2} + e^{-1} - 2)$$

$$= (1 - e^{-1/2} - e^{-1})(a\cos(a/2) - 2\sin(a/2) - 3a^2 \sin(a/2)).$$

From this it is immediate that $\tan(a/2) = a/(2 + 3a^2)$. Now we suppose $A \neq 0$ and divide $f$ through by $A$. Then $f'(1/2) = a$ and from (A-1)

$$\lambda f'(1/2) = \frac{a}{1 + a^2}$$

$$= -\frac{1}{2} \int_0^1 e^{-|y-1/2|} H_y(1/2) f(y)\, dy$$

$$= -\frac{1}{2} \int_0^1 e^{-|y-1/2|} H_y(1/2) \sin(a(y - 1/2))\, dy$$

$$= -\frac{e^{-1/2}}{1 + a^2}(a\cos(a/2) + \sin(a/2)) + \frac{a}{1 + a^2}.$$

In particular, $a \cot(a/2) = -1$.

The solutions of $\tan(a/2) = a/(2+3a^2)$ are approximately $2k\pi$ for integers $k$ and the solutions of $a\cot(a/2) = -1$ are approximately $(2k+1)\pi$. Lemma A.2 makes this precise. Since they do not have any common solutions, $A = 0$ if and only if $B \neq 0$. This completes the argument that Theorem A.2 lists all the eigenfunctions of $K$ with positive eigenvalues.

LEMMA A.2. 1. *The positive solutions of* $\tan(a/2) = a/(2 + 3a^2)$ *lie in the set*

$$\bigcup_{k=1}^{\infty} (2k\pi, 2k\pi + 1/3k\pi),$$

*with exactly one solution per interval. Furthermore, $a$ is a solution if and only if $-a$ is a solution.*

2. *The positive solutions of $a \cot(a/2) = -1$ lie in the set*

$$\bigcup_{k=0}^{\infty} ((2k+1)\pi, (2k+1)\pi + 1/(k\pi + \pi/2)),$$

*with exactly one solution per interval. Furthermore, $a$ is a solution if and only if $-a$ is a solution.*

PROOF. Let $f(\theta) = \tan(\theta/2) - \theta/(2 + 3\theta^2)$. Then $f$ is an odd function, so $a$ is a solution to $f(\theta) = 0$ if and only if $-a$ is a solution. Now,

$$f'(\theta) = \frac{1}{2}\sec^2(\theta/2) + \frac{3\theta^2 - 2}{(3\theta^2 + 2)^2}$$

and so $f(\theta)$ is increasing for $\theta \geq \sqrt{2/3}$. Recall the power series expansion of $\tan\theta$ for $|\theta| < \pi/2$ is

$$\tan\theta = \theta + \theta^3/3 + 2\theta^5/15 + 17\theta^7/315 + \cdots.$$

In particular, for $0 \leq \theta < \pi/2$, $\tan\theta \geq \theta$. Consequently, for $\theta \in (0, \pi/2)$,

$$f(\theta) \geq \frac{\theta}{2} - \frac{\theta}{2 + 3\theta^2} > 0.$$

So $f$ has no roots in $(0, \pi/2)$, and is increasing in the domain in which we are interested. Furthermore, for $k \geq 1$,

$$f(2k\pi) < 0 < +\infty = \lim_{\theta \to (2k+1)\pi^-} f(\theta).$$

The third and fourth quadrants have no solutions since $f(x) < 0$ in those regions. This shows that the solutions to $f(\theta) = 0$ lie in the intervals

$$\bigcup_{k=1}^{\infty} (2k\pi, 2k\pi + \pi),$$

with exactly one solution per interval. Finally, for $k \in \mathbb{Z}_{\geq 1}$,

$$f(2k\pi + 1/(3k\pi)) \geq \tan(k\pi + 1/(6k\pi)) - \frac{1}{6k\pi}$$

$$= \tan(1/(6k\pi)) - \frac{1}{6k\pi}$$

$$\geq 0,$$

which gives the result.

To prove the second statement of the lemma, set $g(\theta) = \theta \cot(\theta/2)$. Then $g$ is even, so $g(a) = -1$ if and only if $g(-a) = -1$. Since $g'(\theta) = \cot(\theta/2) - (\theta/2)\csc^2(\theta/2)$, $g(\theta)$ is negative and decreasing in third and fourth quadrants (assuming $\theta \geq 0$) and furthermore,

$$g((2k+1)\pi) = 0 > -1 > -\infty = \lim_{\theta \to 2(k+1)\pi^-} g(\theta).$$

The first and second quadrants have no solutions since $g(x) \geq 0$ in those regions. This shows that the solutions to $g(x) = -1$ lie in the intervals

$$\bigcup_{k=0}^{\infty} ((2k+1)\pi, (2k+1)\pi + \pi),$$

with exactly one solution per interval. Finally, for $k \in \mathbb{Z}_{\geq 0}$,

$$g((2k+1)\pi + 1/(k\pi + \pi/2))$$
$$= ((2k+1)\pi + 1/(k\pi + \pi/2)) \cot(k\pi + \pi/2 + 1/(2k\pi + \pi))$$
$$= ((2k+1)\pi + 1/(k\pi + \pi/2)) \cot(k\pi + \pi/2 + 1/(2k\pi + \pi))$$
$$= ((2k+1)\pi + 1/(k\pi + \pi/2)) \cot(\pi/2 + 1/(2k\pi + \pi))$$
$$= -((2k+1)\pi + 1/(k\pi + \pi/2)) \tan(1/(2k\pi + \pi))$$
$$< -1,$$

which completes the proof.  $\square$

The exact eigenfunctions for the continuous kernel yield approximate eigenfunctions and eigenvalues for the discrete case. Here we give the proof of Theorem 3.1.

PROOF OF THEOREM 3.1.    That $f$ and $g$ are approximate eigenfunctions for the discrete matrix follows directly from Theorem A.2. Suppose $K$ is the continuous kernel. Then,

$$S_n f_{n,a}(x_i) = \sum_{j=1}^{n} S_n(x_i, x_j)[\cos(a(j/n - 1/2)) - (2/a)\sin(a/2)]$$
$$= \int_0^1 K(x_i, y)[\cos(a(y - 1/2)) - (2/a)\sin(a/2)]\,dy + R_{f,n}$$
$$= \frac{1}{1 + a^2} f_{n,a}(x_i) + R_{f,n},$$

where the error term satisfies

$$|R_{f,n}| \leq \frac{M}{2n} \qquad \text{for } M \geq \sup_{0 \leq x \leq 1} \left| \frac{d}{dx} K(x_i, y)[\cos(a(y - 1/2)) - (2/a)\sin(a/2)] \right|$$

by the standard right-hand rule error bound. In particular, we can take $M = a + 4$ independent of $j$, from which the result for $f_{n,a}$ follows. The case of $g_{n,k}$ is analogous. $\square$

The version of this theorem for uncentered matrices is as follows:

THEOREM A.3. *For $1 \leq i, j \leq n$, consider the matrices defined by*

$$A_n(i,j) = \frac{1}{2n} e^{-|i-j|/n} \quad and \quad S_n(i,j) = A_n - \frac{1}{2n} \mathbf{1}\mathbf{1}^T.$$

1. *Set $f_{n,a}(x_i) = \cos(a(i/n - 1/2))$, where $a$ is a positive solution to $a \tan(a/2) = 1$.*
   *Then*

   $$A_n f_{n,a}(x_i) = \frac{1}{1+a^2} f_{n,a}(x_i) + R_{f,n} \qquad where \ |R_{f,n}| \leq \frac{a+1}{2n}.$$

2. *Set $g_{n,a}(x_i) = \sin(a(i/n - 1/2))$, where $a$ is a positive solution to $a \cot(a/2) = -1$.*
   *Then*

   $$S_n g_{n,a}(x_i) = \frac{1}{1+a^2} g_{n,a}(x_i) + R_{g,n} \qquad where \ |R_{g,n}| \leq \frac{a+1}{2n}.$$

*That is, $f_{n,a}$ and $g_{n,a}$ are approximate eigenfunctions of $A_n$ and $S_n$.*

The proof of Theorem A.3 is analogous to Theorem 3.1 by way of Lemma A.1 and so is omitted here.

PROOF OF THEOREM 3.2. Let $\tilde{f}_{n,a} = f_{n,a}/\|f_{n,a}\|_2$. Then by Theorem 3.1,

$$\left| K_n \tilde{f}_{n,a}(x_i) - \frac{1}{1+a^2} \tilde{f}_{n,a}(x_i) \right| \leq \frac{a+4}{2n\|f_{n,a}\|_2}$$

and, consequently,

$$\left\| K_n \tilde{f}_{n,a}(x_i) - \frac{1}{1+a^2} \tilde{f}_{n,a}(x_i) \right\|_2 \leq \frac{a+4}{2\sqrt{n}\|f_{n,a}\|_2}.$$

By Lemma A.2, $a$ lies in one of the intervals $(2k\pi, 2k\pi + 1/3k\pi)$ for $k \geq 1$. Then

$$
\begin{aligned}
|f_{n,a}(x_n)| &= |\cos(a/2) - (2/a)\sin(a/2)| \\
&\geq \cos(1/6\pi) - 1/\pi \\
&\geq 1/2.
\end{aligned}
$$

Consequently,

$$\|f_{n,a}\|_2 \geq |f_{n,a}(x_n)| \geq 1/2$$

and so the first statement of the result follows from Theorem A.1. The second statement is analogous. $\square$

**Acknowledgments.** We thank Harold Widom, Richard Montgomery, Beresford Parlett, Jan de Leeuw and Doug Rivers for bibliographical pointers and helpful conversations. Cajo ter Braak did a wonderful job educating us as well as pointing out typos and mistakes in an earlier draft.

## SUPPLEMENTARY MATERIAL

**Supplementary files for "Horseshoes in multidimensional scaling and local kernel methods"** (DOI: 10.1214/08-AOAS165SUPP; .tar). This directory [Diaconis, Goel and Holmes (2008)] contains both the matlab (mds_analysis.m) and R files (mdsanalysis.r) and the original data(voting_record2005.txt,voting _record_description.txt, house_members_description.txt,house_members2005. txt,house_party2005.txt) as well as the transformed data (reduced_voting_ record2005.txt,reduced_house_party2005.txt).

S. HOLMES
P. DIACONIS
DEPARTMENT OF STATISTICS
STANFORD UNIVERSITY
STANFORD, CALIFORNIA 94305
USA
URL: http://www-stat.stanford.edu/~susan/
E-MAIL: susan@stat.stanford.edu

S. GOEL
YAHOO! RESEARCH
111 W. 40TH STREET, 17TH FLOOR
NEW YORK, NEW YORK 10025
USA
E-MAIL: goel@yahoo-inc.com
URL: http://www-rcf.usc.edu/~sharadg/